\documentclass[aps,prd,preprint,showpacs,amsmath,amssymb]{revtex4}
\usepackage{epsfig}

\newlength{\www}

\newcommand{\be}{\begin{equation}}
\newcommand{\ee}{\end{equation}}
\newcommand{\ba}{\begin{eqnarray}}
\newcommand{\ea}{\end{eqnarray}}

\newcommand{\bq}{\begin{equation}}
\newcommand{\eq}{\end{equation}}
\newcommand{\bqa}{\begin{eqnarray}}
\newcommand{\eqa}{\end{eqnarray}}
\newcommand{\ben}{\begin{enumerate}}
\newcommand{\een}{\end{enumerate}}
\newcommand{\bc}{\begin{center}}
\newcommand{\ec}{\end{center}}
\newcommand{\bqb}{\begin{eqnarray*}}
\newcommand{\eqb}{\end{eqnarray*}}

\begin{document}

\preprint{PTA/06-20}

\title{\vspace{1cm}
The relevance of virtual electroweak effects in the overall $t$-channel 
single top production at LHC
}

\author{M. Beccaria$^{a,b}$, G. Macorini$^{c, d}$
F.M. Renard$^e$ and C. Verzegnassi$^{c, d}$ \\
\vspace{0.4cm}
}

\affiliation{\small
$^a$Dipartimento di Fisica, Universit\`a di
Lecce \\
Via Arnesano, 73100 Lecce, Italy.\\
\vspace{0.2cm}
$^b$INFN, Sezione di Lecce\\
\vspace{0.2cm}
$^c$
Dipartimento di Fisica Teorica, Universit\`a di Trieste, \\
Strada Costiera
 14, Miramare (Trieste) \\
\vspace{0.2cm}
$^d$ INFN, Sezione di Trieste\\
$^e$ Laboratoire de Physique Th\'{e}orique et Astroparticules,
UMR 5207\\
Universit\'{e} Montpellier II,
 F-34095 Montpellier Cedex 5.\hspace{2.2cm}\\
\vspace{0.2cm}
}

\begin{abstract}
We compute the complete one loop electroweak effects in the MSSM for the eight processes of single top (and single antitop) production
in the $t$-channel at hadron colliders, generalizing a previous analysis performed for the dominant $dt$ final state. 
The results are quite similar for all processes, showing an impressively large Standard Model effect and a generally modest 
genuine SUSY contribution in the mSUGRA scenario. The one loop effect on the total rate is shown and the possibility of measuring it at LHC is discussed.
\end{abstract}


\maketitle

\section{Introduction}
\label{sec:intro}

The relevance of a precise measurement of single top production at hadron colliders 
was already stressed  in several papers in the recent years
~\cite{Tait:1999cf}. 
A well known peculiarity of the process is actually the fact that it offers the unique possibility of a direct
measurement of the $tbW$  ($V_{tb}$)  coupling, thus allowing severe tests of the conventionally assumed properties of the CKM matrix in
 the Minimal Standard Model, and for a very accurate
review of the topics we defer to a very recent article~\cite{Alwall:2006bx}.

For the specific purpose of a "precise" determination of $V_{tb}$, two independent requests must be met.
The first one is that of a correspondingly "precise" experimental  measurement of the process. For
the $t$-channel case on which we shall concentrate in this paper a very recent CMS
study~\cite{Pumplin:2002vw} concludes that,with $10$ fb$^{-1}$ of 
integrated luminosity, one could be able to reduce the (mostly systematic) experimental uncertainty of the cross section below the ten percent
level (worse uncertainties are expected for the two other processes, the $s$-channel and the associated
$Wt$ production, whose cross section is  definitely smaller than that of the $t$-channel). The second request is that of a similarly 
accurate theoretical prediction of the observables of the process. In
this respect, one must make the precise statement that, in order to cope with the
goal of measuring $V_{tb}$ at the few (five) percent level, 
a complete NLO calculation is requested. In the SM this has been
done for the QCD component of the $t$-channel, resulting in a relatively small (few percent) effect~\cite{Stelzer:1997ns}. 
The electroweak effects have been computed very recently at the
complete one loop level within the MSSM for the dominant $dt$ component of the process
(to be defined in Section 2)~\cite{Beccaria:2006ir}.
The main conclusion of that analysis was that, within the SM framework, the
electroweak effect at one loop is sizable in the cross section, of the ten
percent size or even more. This effect is larger than that of the
 QCD component, and would be essential for a genuine
precision test of electroweak physics to be performed at LHC. As stressed in ~\cite{Beccaria:2006ir}, the genuine
SUSY effect would be modest (at the relative two-three percent level) for a class of considered benchmark points in the mSUGRA scenario, 
although possible special effects (typically of threshold
kind) might be visible for particular light sparticles (e.e.stop)/ scenarios.

As we anticipated, from a numerical point of view the $t$-channel cross section is by far the largest
in the single top production processes. In~\cite{Beccaria:2006ir}, we only considered one of the eight possible final states of the process
 (conventionally, four single top and four single antitop cases, as shown in
 Section 2) i.e. the $dt$ state that gives more than fifty percent of the overall
cross section. The aim of this paper is that of generalizing the analysis of~\cite{Beccaria:2006ir} to the remaining
seven final states. This will be done in full analogy with~\cite{Beccaria:2006ir}, computing the complete one loop
electroweak effect in the MSSM. In this way, a full description of the overall single top production
process in the dominant $t$-channel will be available for the electroweak sector,
and concrete conclusions will be available concerning the possibility of performing meaningful tests of the
structure of the CKM matrix, or of possible extensions of the Standard Model, at LHC.

Technically speaking, this paper is organized as follows. In Section 2 the definition of the eight
considered processes will be given, and the main tasks that were fulfilled to perform a complete
one loop electroweak calculation will be indicated. Since the problems to be solved were practically
identical with those already met in~\cite{Beccaria:2006ir}, our description will be whenever possible quick and
essential. In Section 3 we shall define the considered observables and show the results of our
calculation, particular emphasis being given to the value of the total $t$-channel cross section.
Short comments, future plans and conclusions will be given in the final Section 4.

\section{The $t$-channel processes at one electroweak loop}

\begin{figure}[tb]
\centering
\epsfig{file=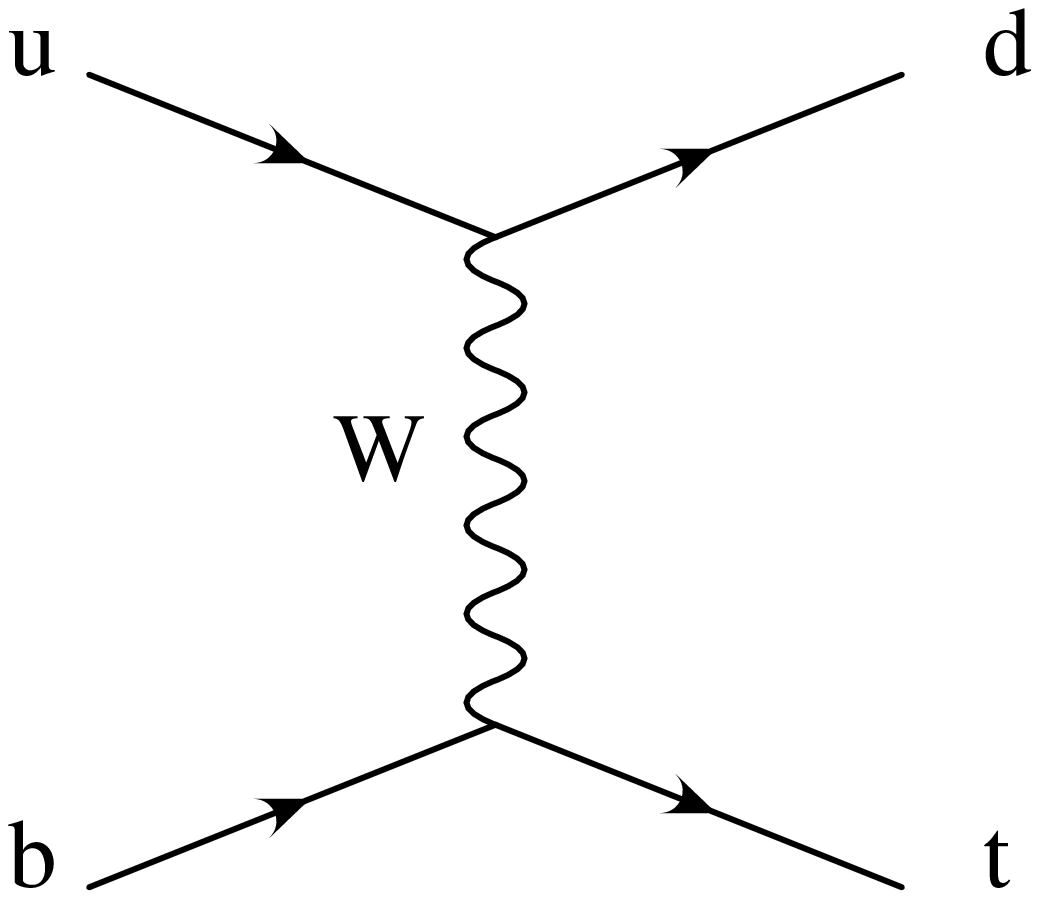, width=6cm, angle=0}
\ \ \ \ \ \ \ \ \ \ \ \ \ \ \ \ 
\epsfig{file=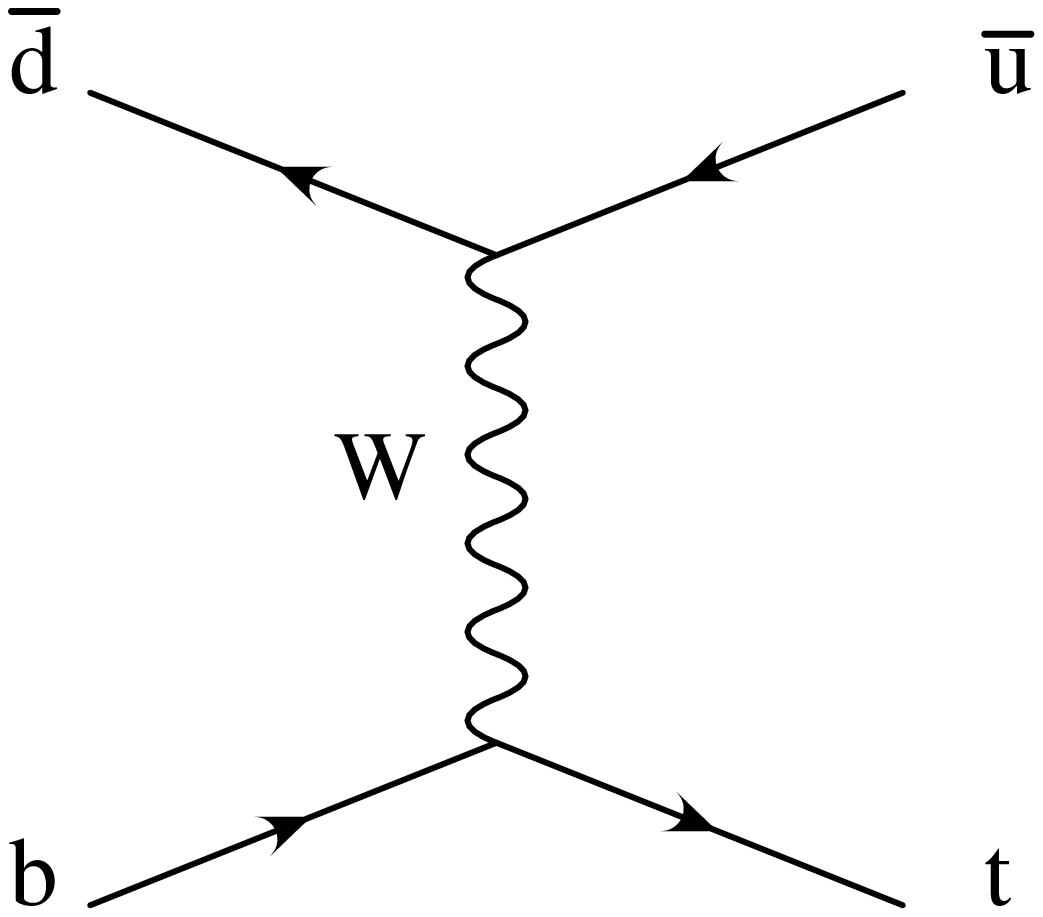, width=6.4cm, angle=0}
\vspace{1.5cm}
\caption{Born direct and crossed processes for single top production in the $t$ channel
with first generation light quark current.}
\label{fig:1}
\end{figure}

The complete description of the $t$-channel  involves at
partonic level four  sub-processes for single $t$ production:  $ub\to td$, 
$\bar{d}b\to t\bar{u}$,  $cb\to ts$,  
$\bar{s}b\to t \bar{c}$,  and the related four for the single
$\bar{t}$ production. \\
The starting point is the cross section for the $ub\to td$ process
with the complete set of one loop electroweak corrections
(in the MSSM and SM) as computed in~\cite{Beccaria:2006ir}, where the reader can find 
 all the relevant formulae and  the technical details.\\
From the results of~\cite{Beccaria:2006ir}, the one loop cross
sections for all the eight partonic processes can be obtained in a straightforward
way, by using a
"crossing" prescription, and replacing the masses.\\
{}\\
At Born level $ub\to td$ is described by a single 
diagram describing $W$ propagation in the $t$-channel and shown in the left part
of Fig.~(\ref{fig:1}).
We define the Mandelstam variables for  $ub\to td$:

\begin{eqnarray}
s&=&(p_b+p_u)^2\nonumber\\
t &=& (p_b-p_t)^2 = m_b^2 + m_t^2 -2 E_b E_t + 2 p p' \cos\theta \nonumber\\
u &=& (p_b-p_d)^2 = m_b^2 + m_d^2 -2 E_b E_d - 2 p p' \cos\theta
\end{eqnarray}

where

\begin{eqnarray}
E_u &=  (s+m_u^2-m_b^2)/(2 \sqrt{s})\qquad
E_b = (s+m_b^2-m_u^2)/(2 \sqrt{s})\nonumber\\
E_t & = (s+m_t^2-m_d^2)/(2 \sqrt{s})\qquad
E_d = (s+m_d^2-m_t^2)/(2 \sqrt{s})\nonumber
\end{eqnarray}
\begin{eqnarray}
p^2   &=  \frac{1}{4s} (s-(m_u+m_b)^2) (s-(m_u-m_b)^2)\nonumber\\
(p')^2 &= \frac{1}{4s} (s-(m_t+m_d)^2) (s-(m_t-m_d)^2)
\end{eqnarray}

and also introduce $q'=p_b-p_t=p_d-p_u$, so $t=q'^2$. The Born amplitude is 

\bqa
A^{Born}=\frac{e^2}{2s^2_W(t-M^2_W)}
[\bar u(t)\gamma^{\mu} P_Lu(b)][\bar u(d)\gamma_{\mu} P_Lu(u)]
\eqa

The "crossed" t-channel
process $\bar d b \to \bar u t$ 
is described by starting from $u b \to d t$
and only exchanging the
$d$ and $u$ lines (leaving the $b$ and $t$ lines unchanged), as shown in the right part of 
Fig.~(\ref{fig:1}). The
incoming $u$ becomes an outgoing $\bar u$, the outgoing $d$ becomes 
an incoming $\bar d$. The scattering angle $\theta$ is still the angle
between $p_t$ and $p_b$ (and now also between $p_{\bar u}$
and $p_{\bar d}$). In terms of Mandelstam variables this is equivalent to exchange
$t \to t, s \to u, u \to s$.\\
The "crossing" rule can be  applied to obtain the unpolarized parton cross
section of $\bar d b \to \bar u t$:

\bqa
 {d\sigma\over d\cos\theta}(\bar d b \to \bar u t)&=&{p'_{cross}\over128\pi sp_{cross}}
\sum_{spins}
|F(\bar d b \to \bar u t)|^2\nonumber\\
&=& {p'_{cross}\over128\pi sp_{cross}}
[\sum_{spins}|F(u b \to d t)|^2](s \to u, u \to s)
\eqa
where $p'_{cross}$ and $p_{cross}$ are obtained from the above $p', p$ by $m_u\leftrightarrow m_d$.

For the $\bar{t}$ production the cross sections can be calculated using the
identities

\bqa
  {d\sigma\over d\cos\theta}{(u b \to t d)} &=& {d\sigma\over
 d\cos\theta}{(\bar{u}  \bar{b} \to \bar{d} \bar{t})}\nonumber\\
 &\phantom{}& \nonumber\\
  {d\sigma\over d\cos\theta}{(\bar{d} b \to t \bar{u})} &=& {d\sigma\over
 d\cos\theta}{(d \bar{b} \to u \bar{t})}
\eqa

and finally the  processes involving the second generation $c,s$ and $\bar{s}, \bar{c}$
quarks can be computed from the previous, simply replacing the masses of the external particles
(and some masses in the loop corrections).\\

\section{Observable quantities and theoretical predictions}

The analysis of this paper will be concentrated on the investigation of
the virtual electroweak            effects on unpolarized cross sections.
The final top polarization can in principle be measured~\cite{Bernreuther:2006pd}
probably not in the first LHC running period, and we shall devote a
forthcoming paper to the study of its main features. Therefore we shall
start with the calculation of the one-loop effects on the inclusive
differential cross sections of the various processes, generally defined
as:
\begin{eqnarray}
\label{eq:basic}
{d\sigma(PP\to t (\bar{t}) q' + X)\over ds}&=&
{1\over S}\sum_{q}~\int^{\cos\theta_{max}}_{\cos\theta_{min}}
d\cos\theta~[L_{q b (\bar{b}) }(\tau, \cos\theta)
{d\sigma_{q b (\bar{b})\to  t  (\bar{t}) q'}\over d\cos\theta}(s)~]
\end{eqnarray}
\noindent
where  the initial quark $q$ and the relative final $q'$ are $q=u,\bar{d},c,\bar{s},$ 
$q'={d,\bar{u},s,\bar{c}}\,$, for $t$ production, and 
$q=\bar{u}, d, \bar{c}, s,$ 
$q'={\bar{d}, u, \bar{s}, c}\,$, for $\bar{t}$ production.
 $\tau={s\over S}$, and $L_{qb}$ is the parton process luminosity,
\be
L_{qb(\bar{b})}(\tau, \cos\theta)=
\int^{\bar y_{max}(tq')}_{\bar y_{min}(tq')}d\bar y~ 
~\left[~ b(x) q({\tau\over x})+q(x)b({\tau\over x})~\right]
\label{Lij}
\ee
\noindent
where S is the total pp c.m. energy, and 
$i(x)$ the distributions of the parton $i$ inside the proton
with a momentum fraction,
$x={\sqrt{s\over S}}~e^{\bar y}$, related to the rapidity
$\bar y$ of the $tq'$ system~\cite{QCDcoll}.
The parton distribution functions are the latest LO MRST (Martin, Roberts, Stirling, Thorne) 
set available on~\cite{lumi}.
The limits of integrations for $\bar y$ depends on the cuts. We have chosen a 
maximal rapidity $Y=2$ and a minimum $p_T$ which we shall specify later.

Note that we are at this stage considering as kinematical observable the 
initial partons c.m. energy $\sqrt{s}$, and not the realistic final 
state invariant mass $M_{tq'}$. 

To relate these two quantities is
relatively straightforward, and we expect from  a previous analysis done
for the top-antitop final state~\cite{Beccaria:2004sx}  that the
difference between them is relatively small. More specifically, we
shall concentrate the discussion of this paper on the properties of
integrated cross sections , the integration being  performed from
threshold to a final realistic energy , so that at this stage the
distinction between c.m. energy and invariant mass looses relevance
although it would be important at a stage where the invariant mass
distribution were actually measured, which again we believe will start
after the first LHC period.
Our starting quantities will therefore be, following the previous
discussion, the inclusive  differential cross sections of the eight
processes, defined in Eq.~(\ref{eq:basic}).
We have computed them in the MSSM model with mSUGRA
symmetry breaking scheme, using a C++ numerical code called LEONE 
available upon request. The one loop amplitudes evaluated inside LEONE
are obtained by crossing symmetry and simple modifications of 
the ones computed in~\cite{Beccaria:2006ir}. As a consequence, they are 
automatically UV finite and reproduce the known asymptotic logarithmic Sudakov expansion,
as we checked in full details in~\cite{Beccaria:2006ir}. Infrared finiteness is less
trivial, but it has also been checked in the present calculation. To this aim,
we have introduced as usual a regulating photon mass $\lambda$ checking that
observables are finite as $\lambda\to 0$.

\begin{figure}[tb]
\centering
\epsfig{file=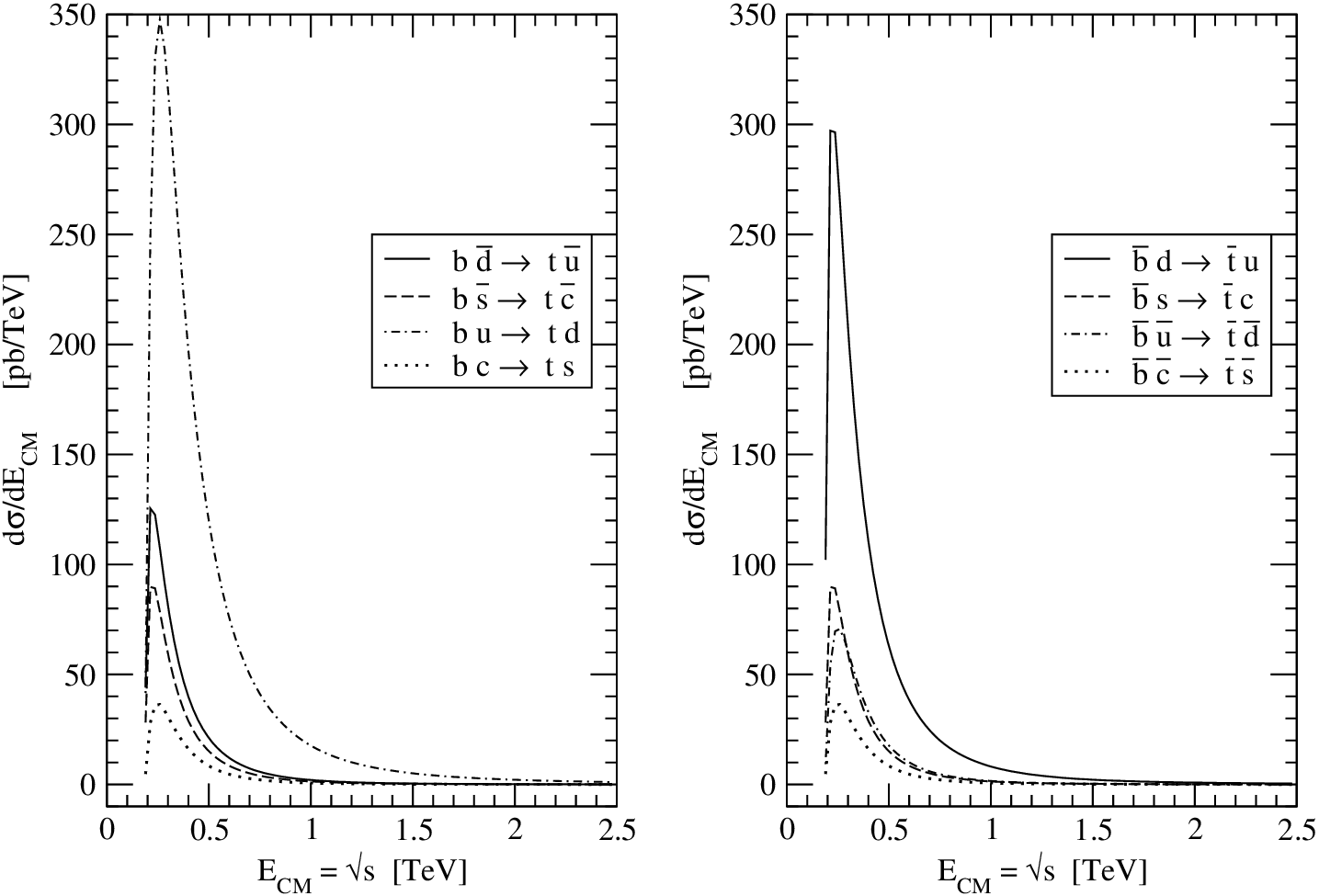, width=18cm, angle=90}
\vspace{1.5cm}
\caption{Differential distribution $d\sigma/dE_{\rm CM}$ for the 4+4 partonic processes
of single $t$  or $\bar t$ quark production.}
\label{fig:2}
\end{figure}

We have analyzed several SUSY benchmark points and,
anticipating a general result, we have systematically found a modest
genuinely supersymmetric ({\em i.e.} beyond the pure Standard  Model) effect.
For this reason we shall only show the results for that benchmark point where
the  small effect is maximum, corresponding to the
ATLAS DC2 SU6 point~\cite{DC2}.
In all the remaining considered points the effect is slightly smaller.
In practice, it remains constantly of the few
percent relative size.
In Fig.~(\ref{fig:2}) we have shown the various distributions for the eight different
processes. One sees that the dominant cross sections correspond to the
final $(t, d)$ and $(\bar t, u)$ pairs, as expected from the corresponding larger
initial states parton distribution functions.  One also notices a generally  smooth shape in energy
for all the computed distributions.
In the next Figures~(\ref{fig:3},\ref{fig:4}) we have shown the total distributions for final
top (\ref{fig:3}) and final antitop (\ref{fig:4}), obtained summing in each Figure the
corresponding four terms of Fig.~(\ref{fig:2}), and separating the Born from the one
loop term, the latter having been computed both for the SM and for the
MSSM. As anticipated, one sees that the difference between the one-loop SM
and MSSM is rather small. To better appreciate it, we have also shown the
relative one-loop electroweak effects. One notices that, while the
relative genuine SUSY effect (i.e. the difference between SM and MSSM)
remains systematically of the few percent size (in particular, two percent at the representative
c.m. energy of 1 TeV), the relative SM effect is impressively large.
Around the 1 TeV energy, chosen for pure indicative reasons , it reaches
the 30 \% size. More relevant for our next analysis, when the
energy is decreased toward threshold, it remains always negative and
sizable, reaching the 10 \% value for low energy values. These
features are valid for both final top and final antitop processes, and we
shall return to this point in the final discussion.

\begin{figure}[tb]
\centering
\epsfig{file=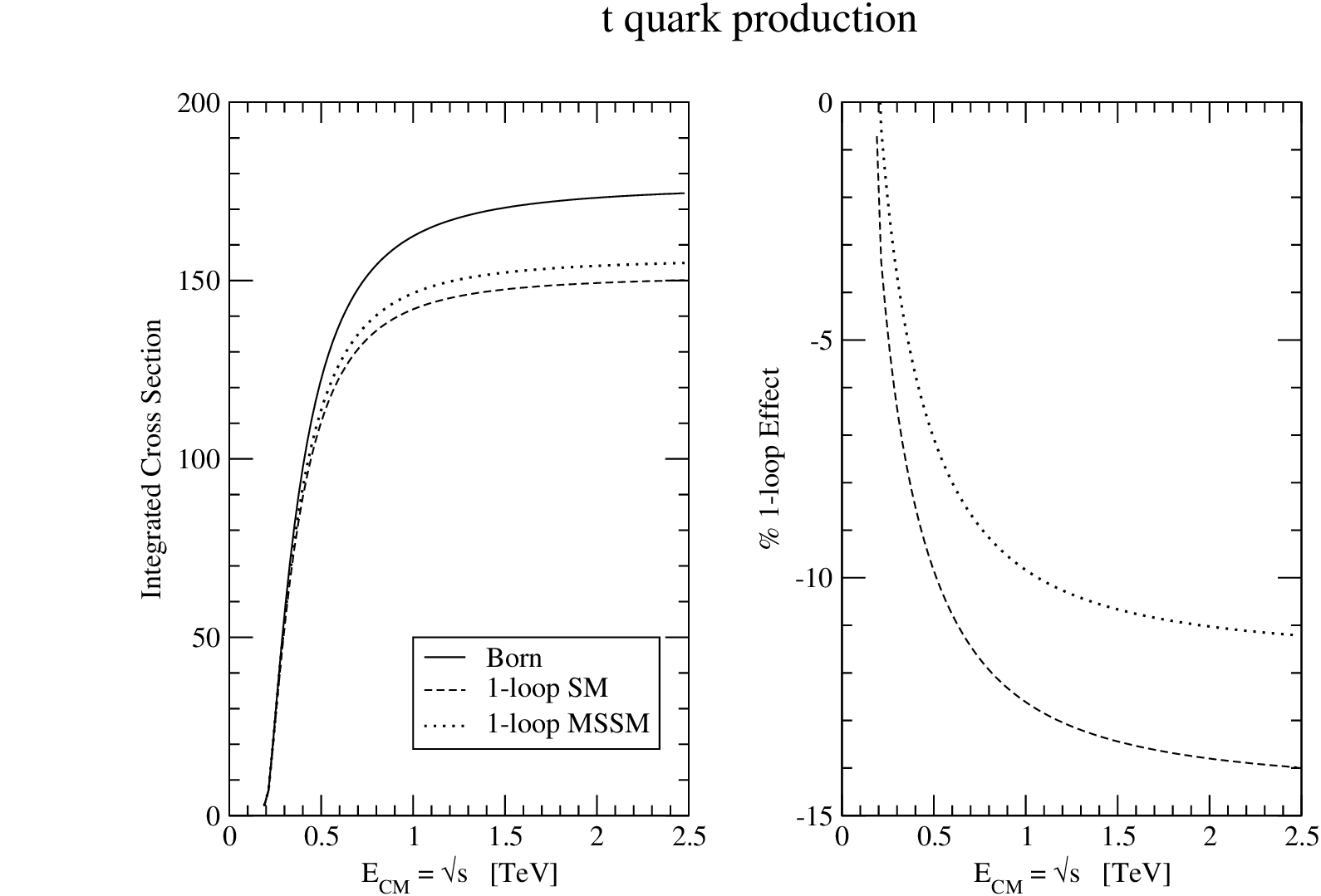, width=18cm, angle=90}
\vspace{1.5cm}
\caption{
Integrated cross section from threshold up to the energy $E_{\rm CM}$ 
for the total rate of production of single $t$ quark at Born level and with full electroweak radiative corrections in the SM and in the MSSM.
The right panel shown the percentual radiative effect in the SM and MSSM.}
\label{fig:5}
\end{figure}

\begin{figure}[tb]
\centering
\epsfig{file=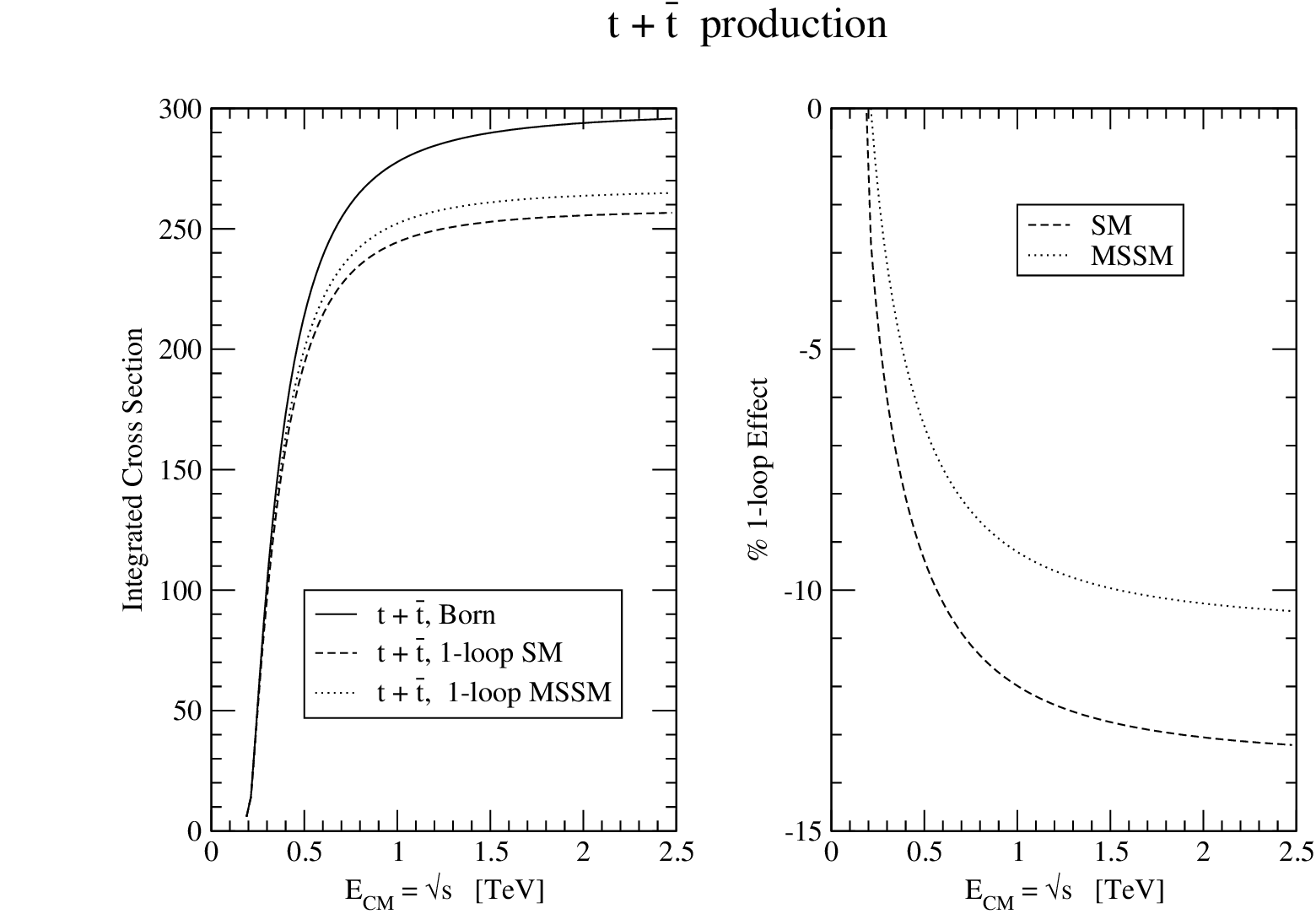, width=18cm, angle=90}
\vspace{1.5cm}
\caption{
Integrated cross section from threshold up to the energy $E_{\rm CM}$ 
for the total rate of production of single $t$ quark or $\bar t$ antiquark at Born level and with full electroweak radiative corrections in the SM and in the MSSM.
The right panel shown the percentual radiative effect in the SM and MSSM.}
\label{fig:6}
\end{figure}

It is likely that in the first period of LHC running a more experimentally
meaningful observable is, rather than the energy distribution, the
integrated cross section. In this spirit, we have shown in Figures~(\ref{fig:5}, \ref{fig:6}) the
result of the energy integration of the previous distributions, performing
the integration from threshold to a variable final energy, to be
fixed by experimental arguments. More precisely, Fig.~(\ref{fig:5}) shows the
integrated cross section for final top, while in Fig.~(\ref{fig:6}) we have shown
the sum of the two integrated cross sections for final top and final antitop,
which might be simpler to analyze in the first LHC period.
The main (and partially unexpected) result of our analysis is, in our
opinion, the fact that the electroweak effect at one loop is large. This
is true for the SM and also, although not as a consequence of sizable
genuine SUSY effects, for the MSSM in the mSUGRA scheme that we have
chosen. One sees that an integration up to the (reasonable) 1 TeV limit
exhibits a SM effect of $\simeq 12$ \%, slightly reduced in
the MSSM. It should be stressed that the available calculations of the
corresponding one-loop QCD effects~\cite{QCDNLO,QCDNLODecay,Frixione:2005vw} produce
definitely smaller results, of the few percent size. This fact will be
commented in the final discussion. At the expected LHC accuracy, the
electroweak effect that we have computed might and should be
experimentally detectable (a more accurate discussion will be the aim of a
forthcoming dedicated paper~\cite{forth2}, and
would allow LHC to perform a rather remarkable  precision test of
electroweak physics.

\section{Conclusions}

As a result of our analysis, we have found that in the process of single
top and single antitop production at LHC a one-loop electroweak effect of
relatively large size on the distributions and on the integrated cross
sections is present in the MSSM, mostly caused by the SM component. There
are two facts that, we believe, deserve a final comment.
The first is the fact that, to our knowledge, the presence of a one loop
electroweak effect that is larger that the corresponding QCD one in a LHC
process only appears for the process of single top production in the t
channel. For this fact we have our personal view, that we present here.
The starting point is that the possibility of finding at one loop an
electroweak effect larger that the corresponding QCD one was already
noticed in a pioneer paper~\cite{Ciafaloni:1998xg} investigating electron-positron
scattering at asymptotic LC energies. The conclusion was that, at very
large energies, the electroweak effect in the hadronic cross section became
larger that the QCD one. This was a consequence of the assumed asymptotic Sudakov
electroweak expansion and of the presence in it of large double
logarithms. This effect was,though, asymptotic. For smaller energies, it
was usually masked by next non leading non logarithmic terms.
Quite generally, the LHC processes that we have examined~\cite{Beccaria:2004sx,Beccaria:2004xk,Beccaria:2006dt,Beccaria:2006ir}
can be decomposed in a sum of different helicity amplitudes. In this set, a
subset of them obeys an asymptotic Sudakov expansion when the energy is
much larger than the one loop involved masses, but another subset is
"reluctant" to behave in this way, typically for the SM when a top quark
with relatively large mass is involved. At asymptotic energies one finds
in fact in the total energy
distribution a Sudakov expansion~\cite{Beccaria:2006dt}, but at low energies this is
overwhelmed by the different values of the "reluctant" amplitudes.
The case of the single top production in the $t$ channel is, in this
respect, so peculiar because in this process only one helicity amplitude
appears. This obeys a Sudakov logarithmic expansion at large energies.
When one decreases the energy, since there do  not exist other "reluctant"
amplitudes, it is not surprising that the smooth logarithmic energy
behavior, in the absence of energy peaks or thresholds, survives until
threshold remaining  not too below the large asymptotic Sudakov logarithmic
limit, which in this case is larger that the QCD one. We cannot claim that
this is a proof of what we observed, but it seems to us, at least,
conceivable.
The second point is the conclusion that the genuine SUSY effect is small. We
insist on the fact that this conclusion was obtained in the considered
mSUGRA symmetry breaking scheme. We cannot exclude that for different
symmetry breaking mechanisms the genuine effect is more sizable. This
question remains open and, in our opinion, it would deserve a special
dedicated rigorous analysis.

\newpage

\newpage
\begin{figure}[tb]
\centering
\epsfig{file=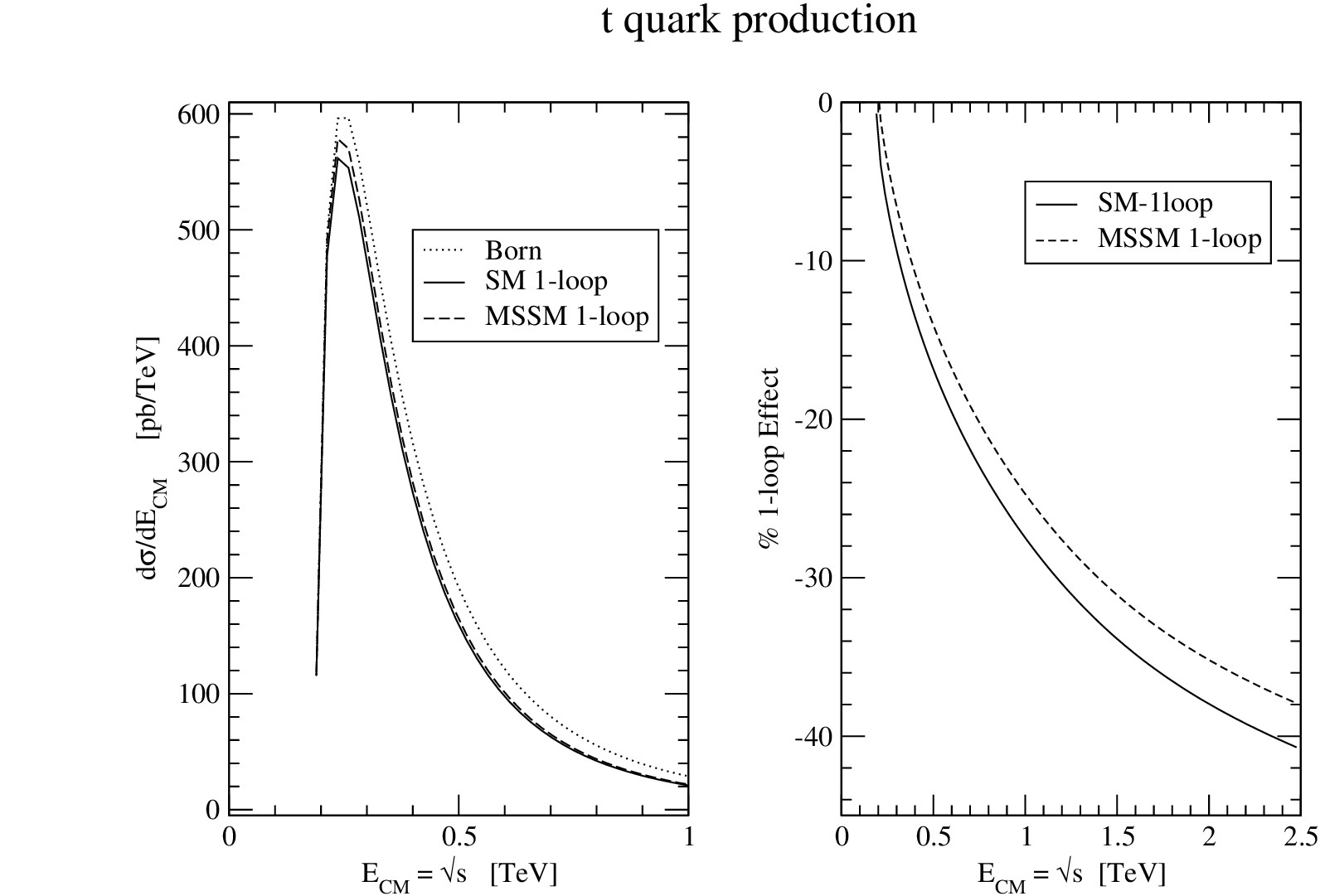, width=18cm, angle=90}
\vspace{1.5cm}
\caption{Differential distribution $d\sigma/dE_{\rm CM}$ for the total rate of production of 
single $t$ quark at Born level and with full electroweak radiative corrections in the SM and in the MSSM.
The right panel shown the percentual radiative effect in the SM and MSSM.}
\label{fig:3}
\end{figure}

\newpage
\begin{figure}[tb]
\centering
\epsfig{file=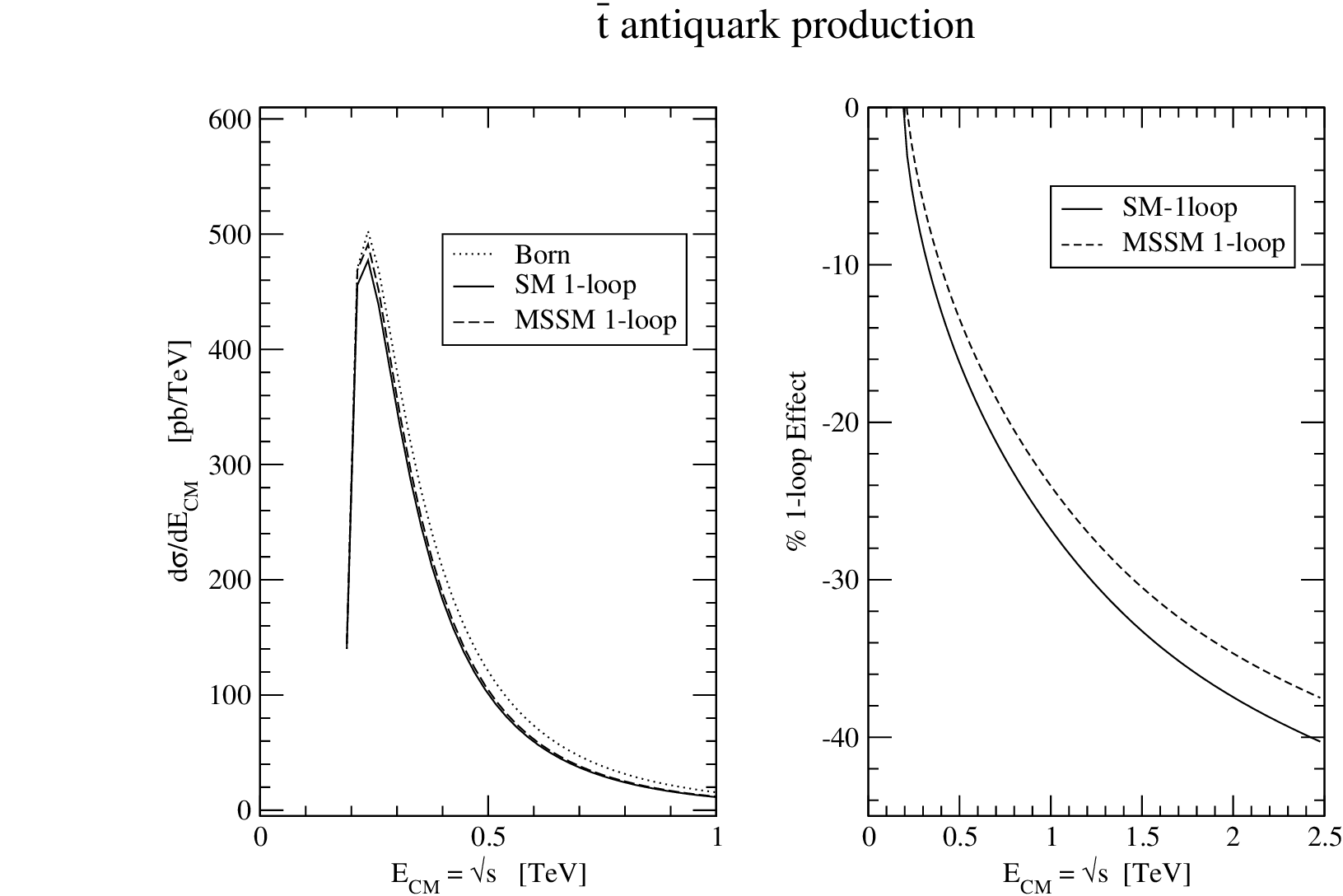, width=18cm, angle=90}
\vspace{1.5cm}
\caption{Differential distribution $d\sigma/dE_{\rm CM}$ for the total rate of production of 
single $\bar t$ antiquark at Born level and with full electroweak radiative corrections in the SM and in the MSSM.
The right panel shown the percentual radiative effect in the SM and MSSM.}
\label{fig:4}
\end{figure}

\newpage

\newpage

\end{document}